\documentclass[12pt, letterpaper, preprint]{aastex631} \usepackage[T1]{fontenc}
\usepackage{color}
\usepackage{amsmath}
\usepackage{natbib}
\usepackage{ctable}
\usepackage{bm}
\usepackage[normalem]{ulem} 
\usepackage{xspace}
\usepackage{csvsimple} 

\usepackage{graphicx}
\usepackage{pgfkeys, pgfsys, pgfcalendar}

\let\oldbibliography\thebibliography 
\renewcommand{\thebibliography}[1]{%
  \oldbibliography{#1}%
  \setlength{\itemsep}{0pt}%
  \setlength{\parsep}{0pt}%
  \setlength{\parskip}{0pt}%
  \setlength{\bibsep}{0ex}
  \raggedright
}
\newcommand{\bitem}{\begin{itemize}}
\newcommand{\eitem}{\end{itemize}}
\newcommand{\beq}{\begin{equation}}
\newcommand{\eeq}{\end{equation}}

\definecolor{orange}{rgb}{1,0.5,0}

\begin{document} \sloppy\sloppypar\frenchspacing

\title{Feature Importance of Climate Vulnerability Indicators with Gradient Boosting across Five Global Cities}

\newcounter{affilcounter}
\author{Lidia Cano Pecharroman}
\altaffiliation{lcano@mit.edu}
\affil{Department of Urban Studies and Planning, Massachusetts Institute of Technology, 77 Massachusetts Ave, Cambridge, MA 02139, USA}

\author{Melissa O. Tier}
\altaffiliation{mtier@princeton.edu}
\affil{School of Public \& International Affairs, Princeton University, 20 Prospect Ave, Princeton, NJ 08540}

\author{Elke U. Weber}
\altaffiliation{eweber@princeton.edu}
\affil{School of Public \& International Affairs, Princeton University, 20 Prospect Ave, Princeton, NJ 08540}

\begin{abstract}
    Efforts are needed to identify and measure both communities’ exposure to climate hazards and the social vulnerabilities that interact with these hazards, but the science of validating hazard vulnerability indicators is still in its infancy. Progress is needed to improve: 1) the selection of variables that are used as proxies to represent hazard vulnerability; 2) the applicability and scale for which these indicators are intended, including their transnational applicability. We administered an international urban survey in Buenos Aires, Argentina; Johannesburg, South Africa; London, United Kingdom; New York City, United States; and Seoul, South Korea in order to collect data on exposure to various types of extreme weather events, socioeconomic characteristics commonly used as proxies for vulnerability (i.e., income, education level, gender, and age), and additional characteristics not often included in existing composite indices (i.e., queer identity, disability identity, non-dominant primary language, and self-perceptions of both discrimination and vulnerability to flood risk). We then use feature importance analysis with gradient-boosted decision trees to measure the importance that these variables have in predicting exposure to various types of extreme weather events. Our results show that non-traditional variables were more relevant to self-reported exposure to extreme weather events than traditionally employed variables such as income or age. Furthermore, differences in variable relevance across different types of hazards and across urban contexts suggest that vulnerability indicators need to be fit to context and should not be used in a one-size-fits-all fashion.

\end{abstract}

\section{Introduction} \label{sec:intro}


As this special issue demonstrates, there is a growing interest among policymakers, community members, and academics in designing and documenting best practices for environmentally-just climate adaptation and resilience policies, including planned relocation. Concern over the need for proactive, equity-based planning of such efforts is justifiably escalating in the face of existential climate crises. Moreover, evidence suggests that policymakers tend to default to status quo policies and decision-making – because of their own cognitive biases~\citep{Roberts2019-lg, Gifford2011-cv}, political pressures they face~\citep{Mayhew2008-ct}, and policy entrenchment~\citep{Wiering2018-ql}; the status quo, in turn, is likely to preserve historical inequities and injustices~\citep{Foster2024-sl}. In particular, community-driven and co-produced planning and research is needed in order to overcome perpetuated social vulnerability and strive toward transformational climate policies~\citep{Foster2024-sl, Morris2024-nh, Ajibade2022-ih}.

To this end, there is a need to better incorporate  vulnerability indicators into climate adaptation policy planning. To start, policymakers and researchers need to better understand \textit{who} is vulnerable to increasing climate hazards – as well as \textit{who} benefits from existing policies and \textit{why} inequalities in both of these domains persist~\citep{Wilson2021-oa}. Recent research finds that key concepts like “vulnerability” and “equity” in flood-risk management are currently operationalized in myriad ways (e.g., as outcomes, units of aggregation, justifications, and more)~\citep{Pollack2024-qi} and also differ depending on whether and which types of "multi-hazards" are considered (e.g., aggregate, cascading, or compound hazards)~\citep{Drakes2022-mi}. Additionally, salient components likely vary location to location, and country to country ~\citep{Rufat2019-ra}. All of these variations clearly play an important role when designing climate adaptation interventions – such as managed retreat/planned relocation – where policymakers have to make value-laden distributional and procedural justice decisions about program design and participants' eligibility~\citep{Siders2022-iq}.

The science of validating vulnerability indicators – let alone translating them into policy approaches – is also in its infancy~\citep{NASEM2024-il}. A recent National Academies report~\citep{NASEM2024-il} distinguishes disproportionate exposure to environmental hazard from other forms of social vulnerability caused by “combinations of political, economic, social, and institutional processes” (p. 25) that also interact with environmental hazard outcomes – as well as from cumulative impacts of exposure. There are many methods for measuring these hazard exposures, cumulative impacts of exposure, and other relevant social vulnerabilities. Such indicators (e.g., the widely used Social Vulnerability Index) as well as composite sets of indicators (e.g., the White House Council on Environmental Quality’s new Climate and Economic Justice Screening Tool or the Environmental Defense Fund's U.S. Climate Vulnerability Index) can then be used as tools to advance our understanding of patterns of injustice and to determine community eligibility for government resources~\citep{NASEM2024-il, Lewis2023-pn}. As such they should be scrutinized and used with caution.

The NASEM report also stresses the importance of community engagement in the design of composite indicators, which inherently involves subjective decision-making processes. In addition to community-led or co-produced processes, social science research methods can help to elicit community input on vulnerability indicators, such as via online surveys, in-person interviews, and focus groups~\citep{Small2022-ju, Groves2009-qy}. Incorporating community input can also help to identify location-specific vulnerability profiles and avoid one-size-fits-all policy approaches. Existing tools do not meet this need. For example, the Climate and Economic Justice Screening Tool (CEJST) notably does not include race or ethnicity, age, disability, or other demographic characteristics that have frequently been associated with differential outcomes to climate risk and a profusion of other environmental injustices (see~\cite{Bolin2018-eh},~\cite{Sultana2021-lg} and~\cite{Wilson2021-oa} for several helpful overviews); and neither CEJST nor the Climate Vulnerability Index (CVI) incorporate community input on the selection of vulnerability indicators.\footnote{Mullen et al. (2023) specifically consider CEJST from the perspective of Indigenous communities and offer several recommendations for better designing and contextualizing indicators. Importantly, they note: "Indigenous peoples should be engaged at every level of the development and implementation of screening tools, in ways that affirm Indigenous self-determination and the principle of free, prior, and informed consent" (p. 367). Such principles hold for other local contexts as well.}

Furthermore, CEJST is a binary screening tool: census tracts are recognized as "disadvantaged" if they meet the threshold for at least one of eight categories of burden (climate change, energy, health, housing, legacy pollution, transportation, water and wastewater, workforce development) or if they are located within the border of a Federally Recognized Tribe or Alaska Native Village. By contrast, CVI – which was created by the Environmental Defense Fund in partnership with researchers at Texas A\&M University and Darkhorse Analytics – includes four types of social vulnerabilities (health, social and economic, infrastructure, and environment) and three types of risks exacerbated by climate change (extreme events, social and economic, and health), each with their own set of holistic sub-categories. Census tracts are assigned a "national vulnerability percentile" by aggregating data across indicators. Although CVI incorporates a much wider range of social vulnerability categories and more climate hazard types than does CEJST, it still only creates one vulnerability profile for any census tract no matter which risks that community faces.

In response to gaps in research and praxis, this paper contributes to the crafting and validating of climate hazard vulnerability indicators that can be employed in policy decision making. In particular, it fills two knowledge gaps: 1) considering the relevance of indicators that have not been traditionally included in composite indices and that respond to appeals to incorporate community input and lived experienced into the understanding of climate hazard vulnerability; and 2) providing evidence on, as well as limitations to, the translatability of these indicators across hazard types and urban geographies – thereby contributing to comparative urban studies. 

We utilized a subset of elicited participant responses from a large-scale, international survey that examined urban residents’ preferences with regard to various flood-related policies. The online survey was translated and administered in five cities globally: Buenos Aires, Argentina; Johannesburg, South Africa; London, United Kingdom; New York City, United States; and Seoul, South Korea. We analyzed these survey data with a supervised machine learning approach, XGBoost ~\citep{Chen:2016:XST:2939672.2939785}, to identify feature importance of  items in the survey instrument and create a comprehensive index of vulnerability to extreme weather hazards. XGBoost uses training data with multiple features to predict a target variable; in this case, the target variable was survey participants' self-reported experience with extreme weather events. Our analysis measured the importance that key variables – self-perceptions of vulnerability and more objective demographic characteristics – had on extreme weather exposure.
\linebreak


\section{Data} \label{sec:data}
The present study employs data resulting from an international survey of approximately 645  residents per city (total $n$=3224) that gathered data on participants' experiences in urban settings of various extreme weather events, with flood management interventions, of their demographic characteristics, and of their preferences among different policies geared towards reducing flood disaster risk (see Appendix A for the survey codebook). Out of this larger dataset, we utilized variables related to residents' experiences with extreme weather events and their demographic characteristics. Out of the 3224 surveyed, we discarded 823 responses from participants who had lived in their city for less than three years, to remove responses from respondents who may not have had enough time to experience hazards in their city. We also discard 13 responses from participants responding with a non-binary gender identity because these responses only account for 0.4\% of the sample and so do not allow for robust conclusions. In total, we used $n=2,388$ responses in this study.
More specifically, we utilized the following survey items:
\begin{itemize}
    \item \textbf{Exposure to extreme weather events.} Participants rated their exposure to eight extreme weather events on a 5-point Likert-type scale (1 Never; 2 Rarely - has occurred for me once or twice in my life; 3 Sometimes - has occurred for me every 2-5 years; 4 Regularly - has typically occurred for me 1-2 times each year; 5 Frequently - has typically occurred for me 3+ times each year). The eight rated extreme weather events were: flooding from heavy rainfall, flooding from coastal storms, flooding from river overflows, heavy winds, heatwaves/extreme heat, droughts, wildfires, and earthquakes. These hazard types were selected to represent a breadth of extreme weather events exacerbated by climate change. Earthquakes were included because, in pilot testing, participants frequently added this hazard type as a risk event that they had experienced.

    \item \textbf{Demographics.} Demographic questions included: education level, income, age, gender, queer identity (binary yes/no), disability identity (binary yes/no), and whether they speak a non-dominant language as their primary language at home (binary yes/no with a text entry if yes; what was listed as a dominant language varied by city). Education, income, age, and gender are common demographic items in survey instruments and moreover, have received attention in the literature on climate and environmental inequities. We also included queer identity, disability identity, and non-dominant primary language as demographic items in our instrument because they have been understudied as vulnerability indicators. We note that a race/ethnicity item was also included as part of the larger survey instrument; however, we do not include it as a variable in this study. First, categorical variables such as race hold no meaning for the feature importance analysis described below; we cannot translate racial/ethnic categories into numerical measures. Second, even if we were to create a yes/no binary variable of minority racial/ethnic identity, it would be difficult to standardize this measure across our city locations and so a comparative analysis would not be possible.

    \item \textbf{Self-perceptions of vulnerability.} Two survey questions asked about participants' sense of their own vulnerability: 1) an ex-ante assessment of their likely exposure to harm from future flood disasters as compared to other residents of their city and 2) an ex-post assessment of their recent experience with discrimination. Both items were on 5-point Likert-type scales. hese two items were included to distinguish between \textit{actual exposure} to hazard events (albeit self-reported exposure) and participants' evaluation of their own personal risk of harm or societal bias.
   
\end{itemize}

\section{Methods {\sc}} \label{sec:methods}

\subsection{City Selection and Survey Design
} \label{sec:results1}

The international survey was designed through an iterative process that included each of the case study locations. We selected cities based on similarities in population size, national governance type, economic standing, and experience with extreme weather events. We also selected for global geographic range, and made sure that the case locations included multiple cities in the Global South. The survey instrument was designed as follows: 1) it was first drafted in English; 2) collaborators from each city checked for cultural fit of the questions and we edited the English version as needed; 3) collaborators provided translations in conjunction with additional rounds of translation and back-translation via DeepL and Google Translate; \textit{different} colleagues from each city edited the translations as needed and we checked for consistency across all locations and versions. The survey platform was built using Qualtrics XM software, approved by [\textit{name redacted for peer review}] University's Institutional Review Board (\#16462), and beta-tested internationally using Prolific. The actual data collection was conducted via the Centiment survey panel company.

\subsection{Estimation: Feature Importance with XGBoost
} \label{sec:results1}

This study aims to understand which characteristics of the survey population are more strongly associated with an individual participant's exposure to a particular type of extreme weather event. We assess nine participant characteristics, split into three categories: four commonly referenced climate vulnerability demographics (income, age, gender, and education level), three rarely-referenced climate vulnerability demographics (queer identity, disability identity, and non-dominant primary language), and two self-reported perceptions of vulnerability (self-perceived vulnerability to flood risk and self-perceived discrimination).

Feature importance indicates the degree to which these participant characteristics are present in urban populations experiencing each of the extreme weather events included in our survey. In other words, feature importance assigns a weight to different participant characteristics (self-perceived discrimination, income, etc.) based on how relevant they are for predicting a given target variable (exposure to coastal flooding, heatwaves, etc.). 

We employ a supervised machine learning approach whereby importance analysis is performed through gradient-boosted decision trees to measure the importance of each of the participant characteristics relative to each other, for a given climate hazard type. We chose to employ this method over OLS or other regression-based approaches for three reasons. First, in this case it is not appropriate to assume a linear relationship between hazard exposure and participant characteristics. Second, decision trees are a better choice for interpreting feature importance because here they rank participant characteristics based on their contribution to exposure to hazard events. In particular, our decision trees create ranks based on how much each participant characteristic contributes to the decision-making at each split in the tree. Hence, those participant characteristics that are selected for splitting higher up in the tree are interpreted as more relevant. Third, the participant characteristic variables are either ordinal (e.g., Likert-type scales, income range, etc.) or binary; decision trees, being a non-linear model, have the capability to include variables of this nature. It is also worth noting that we use gradient-boosting trees as opposed to random forests. Gradient-boosted trees allow for higher accuracy because they are trained sequentially, rather than independently, to correct on each others' errors.

We performed this analysis through the XGBoost library~\citep{Chen:2016:XST:2939672.2939785}, which allows for implementation of the stochastic gradient-boosting algorithm~\citep{10.1214/aos/1013203451}  with decision trees being used as the weak learner. 

Like other supervised learning models, tree boosting training is done by defining an objective function and optimizing it. An additive model, 

\begin{equation} \label{eq:composite}
    \sum_{t} p_{t}\,h_{t}(X) 
\end{equation} 

is fit by stages, also called additive training. In each stage a weak learner is introduced, minimizing the loss with each added tree. Trees continue to be added as long as their addition improves the model output. Training stops once the loss function reaches an acceptable level, or when it is no longer possible to improve on an external validation dataset. In this case, we employ a validation set for early stopping to find the optimal number of boosting rounds.   

\begin{equation} \label{eq:composite}
F(m) = F(m-1)+n_* - {\frac{\partial\,{\rm Loss~function}}{\partial\,{\rm Previous~model~output}}}
\end{equation}

Where (eq 2) for any step m, ensemble step m equals ensemble step m minus 1 plus the learning rate (n) times the weak learner at step m. 

The model hence continues training until the validation score stops improving; in this case, the objective is to minimize the log loss. In other words, we employ negative log likelihood as our loss function. This measures how well the model predictions align with the data, or in other words its capability to reproduce true data. This loss function calculates the logarithm of the likelihood of observing the data given the parameters of the model, penalizing the model when it assigns low probability to what in the data would be a correct feature. Following existing convention, the data were split for training (80 per cent) and testing (20 per cent). 

XGBoost provides feature importance in the form of gain, cover, and frequency. \textit{Gain} measures the improvement in accuracy brought by a feature to the tree branches it is on. \textit{Cover} looks at the number of samples affected by the feature for each split. \textit{Frequency} is the percentage of times a covariate is used in splitting. In this case, we employ gain in order to understand the contribution of our covariate of interest to the accuracy of the prediction. 

\begin{figure*}
    \centering
    \includegraphics[width=0.8\textwidth]{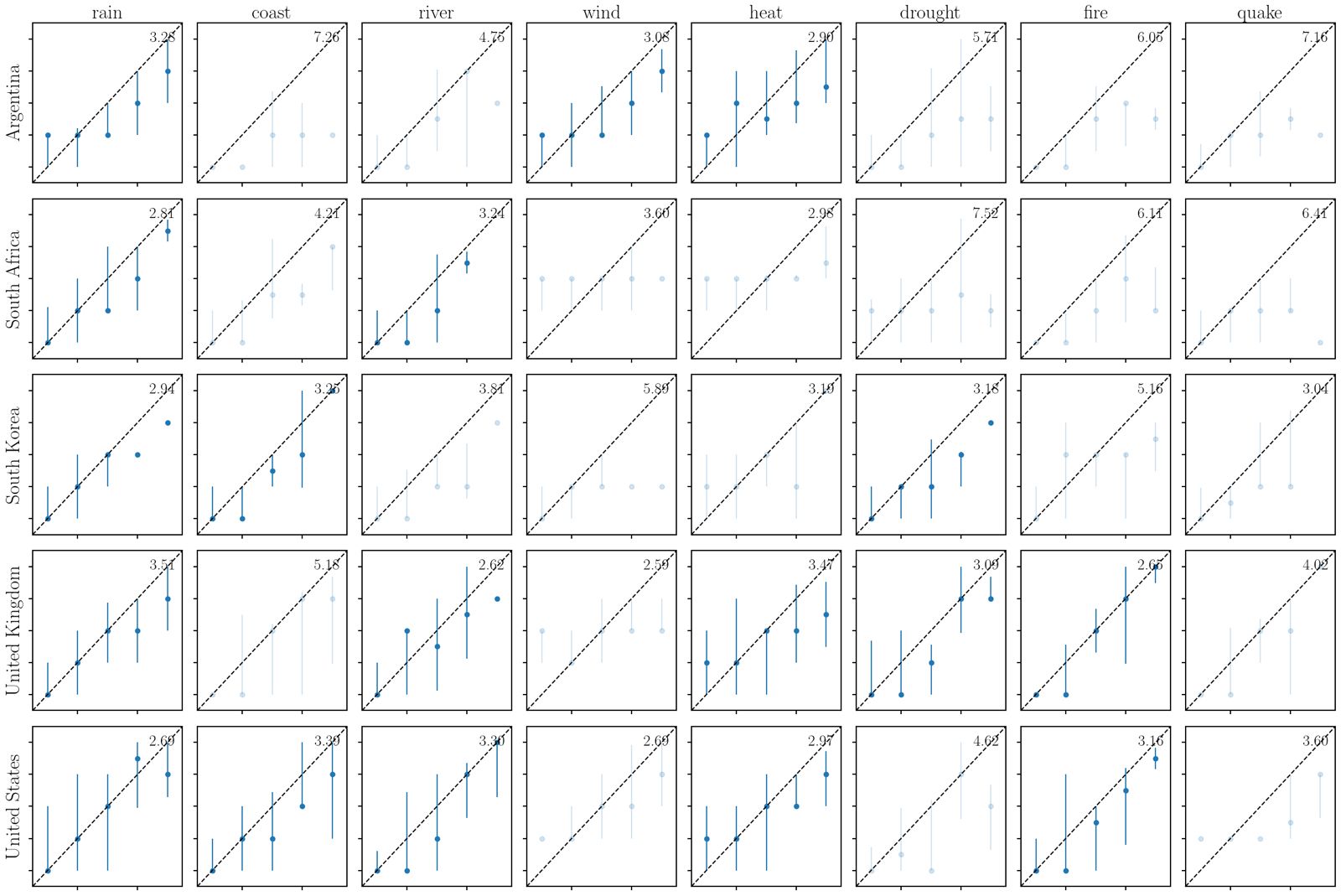}
    \caption{
    Results validation for all cities and across extreme weather hazard types. The numerical figure in the top right corner of each square is the weighted RMSE for the model. Figures that are transparent are conservatively discarded because the fit of the model was not deemed sufficiently robust. 
    This is determined both with visual inspection and by examining the RMSE. The plot shows the performance of the predictive model. The lines are the plotted median predictions of the five best trained models and the uncertainty of such predictions for the different target classes (in this case the 1 to 5 scale that measures each hazard exposure) based on the different participant characteristics of importance. The x-axis is the true score, while the y-axis is the predicted score. The dashed line serves as a reference line where predicted values equal true values.   
    }
    \label{fig:validation}
\end{figure*}

Once models are fitted, the weighted RMSE of the model is used together with visual inspection to determine the appropriate threshold for results. Only models with a robust fit are included in the results. Out of the models represented, results that are not considered are those hazard-city combinations with a high RMSE and poor visual fit. Figure~\ref{fig:validation} shows how the validation of the result was conducted. The series of plots show the performance of the top  trained models, where the dots represent the median prediction and the bar represents the uncertainty of such prediction for each target class. Performance in this case is measured as the model's ability to predict the true value of hazard exposure (represented in the x-axis) through the predictions represented along the y-axis. The predictions represent values from 1 to 5, which is the scale at which exposure was measured. For instance, exposure to heavy wind was expressed on a scale of 1 to 5, and the plot shows how accurate the model is at predicting such exposure given the inputted participant characteristics (income, gender, discrimination, etc.).  The fit to the line represents the fit of the models: the closer to the dotted line, the better the model can predict exposure. The number in the top right corner of each figure is the weighted RMSE for the model. The transparency of some figures indicates that the fit of the model was not sufficiently strong. 

\begin{figure*}
    \centering
    \includegraphics[width=0.8\textwidth]{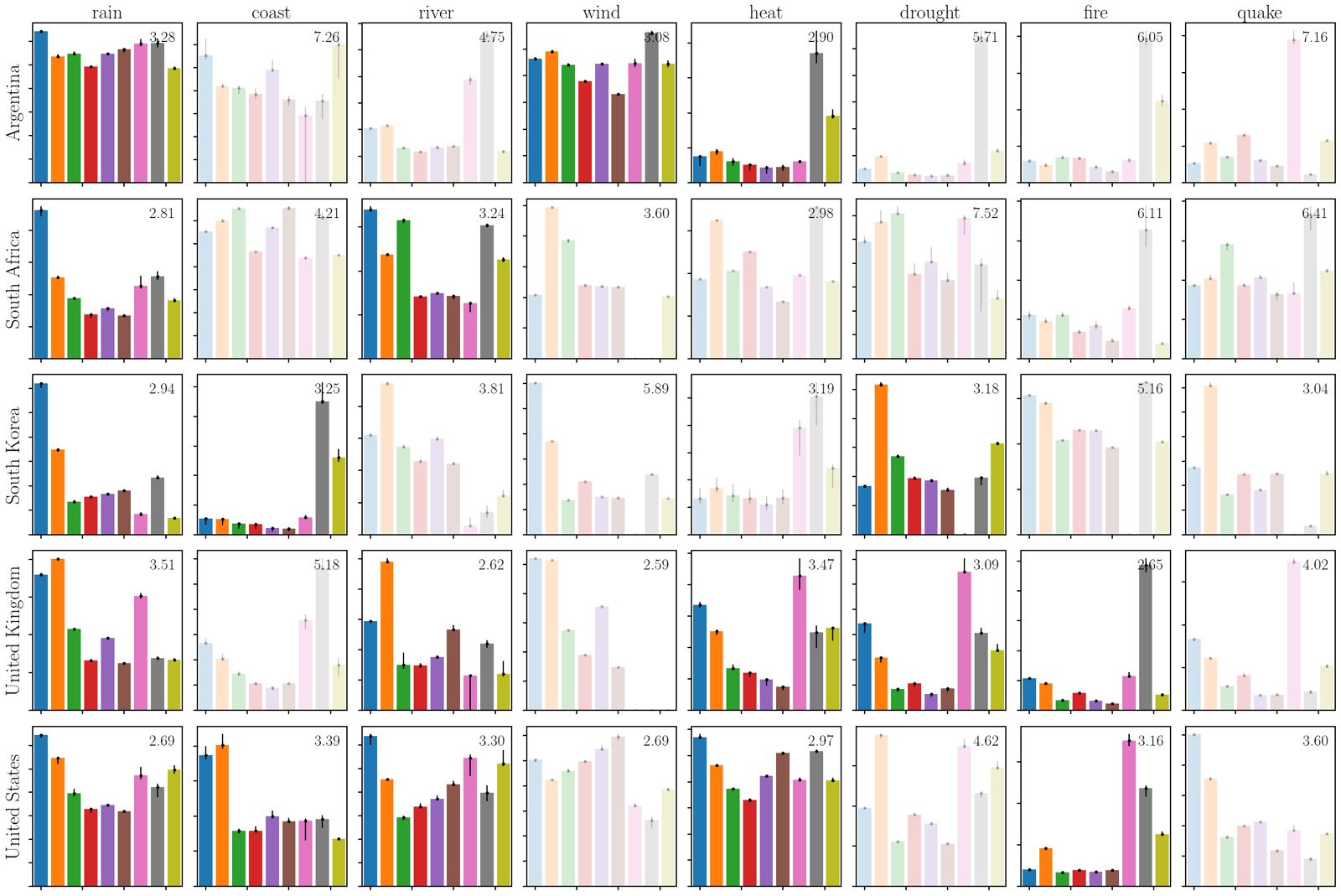}
    \caption{
    Representation of the "feature importance" of participant characteristics. The height of the bar reflects each characteristic's relative importance with respect to the other characteristics for a given extreme weather hazard type and city.  As in Figure~\ref{fig:validation}, the results corresponding to models not deemed sufficiently robust are transparent. 
    The colored bars represent the following participant characteristics: self-perceived harm from flood risk (dark blue), self-perceived discrimination (orange), education level (green), income (red), age (violet), gender (brown), queer identity (pink), disability identity (gray), and non-dominant primary language (yellow). The error bars represent the range of results from the top five models trained. 
    }
    \label{fig:results}
    \centering
\end{figure*}

Figure~\ref{fig:results} shows the primary results of our study. The histogram bars represent the feature importance calculated by the best fitted model. Each bar represents a given characteristic that our model looks at. The height of histogram bars represents the relative importance of a characteristic with respect to the other characteristics for a given city and hazard. The error bar at the top of each histogram bar represents the range in importance when calculated with the top five best fitting models after training. For reference, a total of 1000 models were trained. This is meant to show not only the model chosen as the best fit, but also the range of results if we had picked alternative models that constitute a good fit. This showcases that even though different training rounds can deliver different results, these are highly consistent across resulting models.

\section{Results and Discussion {\sc}} \label{sec:results}

The results show that vulnerability to extreme weather events is complex and that participant characteristics associated with such vulnerability are far from homogeneous. Climate vulnerability indicators and screening tools must consider this heterogeneity in order to achieve their intended goals. Our results first suggest that the type of variables that are most commonly associated with climate vulnerability (such as income, educational attainment, or minoritized identities) are not the only features to keep an eye out for when predicting who is at risk for climate extremes in urban environments. We also show that different characteristics are relevant not only across extreme events but also across urban environments. We discuss each of these findings in further detail, then turn to a policy application in the domain of managed retreat/planned relocation.

\subsection{New Salient Features: Additional Demographics and Self-Perceptions of Vulnerability} \label{sec:results1}

Across the 18 models deemed robust (nontransparent panels in Figure~\ref{fig:results}), some similarities stand out. First, more traditional vulnerability indicators (i.e., education level, income, age, and gender) are less frequent among the most important characteristics for any of the surveyed city populations. Second, some additional demographic variables from our survey were in fact more important than traditional vulnerability indicators (i.e., queer identity, disability identity, and non-dominant primary language). Third, measures of perceived discrimination and perceived vulnerability to flood risk in particular were frequently the most important predictors across different types of extreme weather events and across cities. 

In the Buenos Aires sample, the participant characteristic most important to exposure to flooding from extreme rainfall was self-perceived harm from flood risk and the least important were income and non-dominant primary language (with relatively high characteristic importance across the board). Disability identity was the most important and gender was the least important to exposure to heavy winds (again with relatively high importance across the other characteristics). Disability identity was the most important to exposure to extreme heat, with non-dominant primary language somewhat important and all other characteristics ranked very low.

In the Johannesburg sample, self-perceived harm from flood risk was by far the most important to exposure to flooding from extreme rainfall, while income and gender were the least important (and moderate importance across the other characteristics). Self-perceived harm from flood risk was again the most important to exposure to flooding from river overflows, with education level and disability identity also highly important, self-perceived discrimination and non-dominant primary language moderately important, and all others ranked lower.

In the Seoul sample, self-perceived harm from flood risk was again the most important by a significant amount to exposure to flooding from extreme rainfall, self-perceived discrimination and disability identity were moderately important, and queer identity and non-dominant primary language ranked very low. Disability identity was the most important by a significant amount to exposure to flooding from coastal storms, non-dominant primary language was moderately important, and all other characteristics were ranked very low. Self-perceived discrimination was by far the most important to exposure to droughts, with moderate importance for all other characteristics (queer identity was not included due to absence in the sample).

In the London sample, self-perceived discrimination was most important to exposure to flooding from extreme rainfall, with self-perceived harm from flood risk and queer identity also very important and all other characteristics moderately important. Self-perceived discrimination was also most important to exposure to flooding from river overflows, with all other characteristics moderately important. Queer identity was most important to exposure to extreme heat, with self-perceived harm from flood risk, self-perceived discrimination, disability identity, and non-dominant primary language all moderately important and remaining characteristics ranked low. Queer identity was most important to exposure to droughts, with self-perceived harm from flood risk, self-perceived discrimination, disability identity, and non-dominant primary language all moderately important and remaining characteristics ranked low. Disability identity was by far the most important to exposure to wildfires, with all other characteristics ranked quite low.

In the New York City sample, self-perceived harm from flood risk was most important to exposure to flooding from extreme rainfall, with self-perceived discrimination as next-most important and all characteristics having moderate to high importance. Self-perceived discrimination was most important to exposure to flooding from coastal storms, with self-perceived harm from flood risk a close second and all other characteristics moderately important. Self-perceived harm from flood risk was also most important to exposure to flooding from river overflows, with queer identity and non-dominant primary language as next-most important in turn and all other characteristics having moderate to high importance. Self-perceived harm from flood risk was also most important to exposure to extreme heat, with disability identity and gender as next-most important in turn and all other characteristics having moderate to high importance. Finally, queer identity was most important to exposure to wildfires, with disability identity as next-most important, self-perceived discrimination and non-dominant language as moderately important, and all other characteristics ranked quite low.

If we take frequency counts for which participant characteristics were ranked as most important across all 18 robust models, self-perceived harm from flood risk is identified the \textit{most} often (seven times), with self-perceived discrimination and disability identity both identified four times and queer identity three times. Education level, income, age, gender, and non-dominant primary language are never the most important characteristic for any city-hazard pair. Conversely, gender is identified as the \textit{least} important characteristic eight times – with non-dominant primary language identified as the least important three times; education level, age, and queer identity identified two times each, and income identified once. Self-perceived harm from flood risk, self-perceived discrimination, and disability identity are never the least important characteristic. Notably, non-dominant primary language often falls in the middle of the pack in terms of feature importance – more often than do education level, income, age, and gender.

These findings are important to highlight because composite indicators are typically built to measure proxies for risk and vulnerability to hazard events. However, proxies are not always well fit to measure the complex relationships that lie behind them. Importantly, they are also not well suited to capture the lived experience of urban residents, which can be crucial in defining their resilience to exposure events. Our results show the importance of including measurements, either quantitative or qualitative, that can help decision-makers to have a sense of context-specific vulnerability characteristics. Measurements of self-perception in particular are as crucial for composite indices as other proxies of vulnerability. It is important to stress that our survey distinguishes between \textit{actual} exposure to hazard events (albeit self-reported exposure) and participants’ evaluations of their own vulnerabilities. The hazard exposure survey items queried participants about the number of times that they had experienced each hazard type – rather than their exposure in relation to their neighbors or peers. On the other hand, the two vulnerability questions were posed to have participants consider their \textit{relational} vulnerability within some perceived societal average. Thus, it is a consequential finding that participants' perceptions of their own vulnerability does in fact match how their hazard exposure compares to that of others.

\label{sec:results1}

\subsection{Salient Features Influenced by Hazard Types and Location} \label{sec:results2}
Despite the commonalities described above, we also find that there is significant heterogeneity regarding feature importance across cities and across hazard types. For example, wildfires – as well as droughts, extreme heat, and coastal flooding for certain cities – were one hazard type in which self-perceptions of vulnerability were much more subdued while queer and/or disability identity were far more prominent. For the three extreme weather events related to flooding (heavy rainfall, coastal storms, and river overflows), self-perceptions of vulnerability (harm from flood risk and/or discrimination) were more consistently important.

Local context also plays an important role, especially considering pre-existing variation regarding which identities experience marginalization. For example, disability identity repeatedly ranked highly in the Buenos Aires sample – and no other city – as a predictor for the three hazard types represented in this city's robust models (exposure to flooding from heavy rainfall, heavy winds, and extreme heat), despite more variation in the other participant characteristics across hazard types for Buenos Aires. Another example is that queer identity in the Seoul sample shows an unusual pattern compared to other cities: it ranks very low in importance for exposure to flooding from heavy rainfall and coastal storms, and has no data for droughts (i.e., no Seoul participants who identified as queer also had experienced droughts). This could suggest that this participant characteristic is not tied to vulnerability in Seoul (unlike in other cities) – or, perhaps more likely, that participants were much less likely to self-report this identity in this location.

These results are in line with existing literature that highlights the importance of indicator weights (i.e., giving numerical importance to some indicators over others) when employing such vulnerability indicators for disaster decision-making~\citep{Papathoma-Kohle2019-md}, as well as the importance of considering "dynamic vulnerability" to understand disaster risk~\citep{de-Ruiter2022-cd}. Dynamic vulnerability refers to the fact that vulnerability to a given hazard might change over time and can be influenced by changes in context and even by compounded hazards. Our results showcase that the risk of harm from climate hazards needs to be understood within the contextual idiosyncrasies of a given city, and vulnerability indicators cannot be generalized across cities or even across hazards within the same city. This is particularly true when selecting the types of questions or measurements that will be assessed about a population. Furthermore, certain aspects key to vulnerability may not be self-reported due to cultural or contextual reasons, so other approaches may need to be considered to truly gauge relevant features and to support those that are most vulnerable. 
\label{sec:results2}

\subsection{Takeaways in the Context of Managed Retreat} \label{sec:results3}

This article was written in the context of this journal's special issue titled, \textit{Managed Retreat in Response to Climate Hazards}. In this last section we include reflections on planning for managed retreat in order to demonstrate a concrete policy area that could benefit from our findings.

With the "solution space" for managing the global climate crisis shrinking as dangerous effects of greenhouse gas emissions become increasingly "locked-in"~\citep{Haasnoot2021-ls}, ambitious and transformative adaptation planning and implementation is needed now. Adaptation actions are commonly categorized as either \textit{resistance} (such as building a seawall), \textit{advancement} (such as creating new land through reclamation), \textit{accommodation} (such as elevating a structure at risk of flooding), \textit{avoidance} (such as restricting development in high-risk floodplains), or \textit{retreat} (such as planned relocation)~\citep{Mach2021-yb}. Retreat, the focal point of this special issue, typically comes into play when other adaptation measures are deemed insufficient given the level of exposure or vulnerability to the hazards that a given community or household is experiencing. \textit{Managed} retreat refers to instances when relocation is purposeful and coordinated, although it may be initiated over different times and scales using various instruments by actors – including various levels of government, the private sector, and civil society.

Even though effective in reducing hazard exposure, managed retreat faces many barriers to success – and even the precise terminology for this type of policy initiative is highly debated.\footnote{\cite{Baja2021-qy} describes that both “being managed” and “retreat” are often seen as undesirable states, especially in the U.S.; also, “managed”, “retreat”, and “relocation” all have significant connotations in the U.S. – especially for people of color – of past involuntary movements forced by the federal government (including via slavery of Black people, genocide and other removals of Indigenous people, internment of Japanese-Americans, redlining of various minoritized groups, etc.).~\cite{Ajibade2022-ih} also note that while managed/planned retreat is a common term in the Global North, relocation and resettlement are more common terms in the Global South.} In the U.S., the limited number of climate-induced managed retreat programs that do exist have largely been tackled through buyout/acquisition strategies~\citep{Siders2021-qp}. Though there is a growing literature on these buyouts (see~\cite{Greer2022-js} and~\cite{Mach2019-fp} for recent review papers),~\cite{Greer2017-mv} also conclude that minimal policy learning has occurred due to lack of both data and formal government guidance. Evidence consistently shows that buyout programs suffer from a wide range of challenges, including: long wait times~\citep{Siders2021-qp, Weber2019-ra}; inadequate federal funding~\citep{CRS2022-ol, Peterson2020-pl}; complex and uncoordinated multi-level governance processes~\citep{Siders2021-qp, Weber2019-ra}; lack of transparency in program procedures/structures~\citep{Greer2022-js}; excessive focus on individual homeowners, as opposed to renters, communities, or other forms of housing tenure~\citep{Wilson2021-oa}; failure to incorporate social vulnerability indicators~\citep{Wilson2021-oa}; insufficient monitoring and evaluation~\citep{McGhee2020-qs}; and misguided targets, such as measuring number of household recipients instead of favorable long-term outcomes~\citep{Manda2023-tf}.  

Of particular importance to relocation programs is ensuring more distributive, procedural, and other forms of climate justice – especially given that current data suggest poor existing outcomes on these fronts.~\cite{Mach2019-fp} find that wealthier and denser counties are more likely to \textit{implement} buyout programs – but within those counties, households in lower-income and more racially diverse areas are more likely to actually \textit{receive} a buyout.~\cite{Loughran2019-dp} find that “the rate at which minority populations were growing” influences White households’ interest in accepting buyouts.~\cite{Curran-Groome2021-zx} also note immense inefficiencies in application processes, which place a burden on lower-income and less staffed local governments. Finally,~\cite{Marino2018-ru} contends that buyouts as currently formulated do not work well in many Indigenous communities where property ownership is not always documented and individually-focused relocations conflict with community-oriented decision-making. 

The most prevalent typology of managed retreat globally is still what~\cite{Ajibade2022-ih} describe as "techno-managerial", or focused on simple hazard exposure reduction tactics. Present-day interventions lag behind not only in overcoming barriers to make the retreat process happen, but also in focusing on the outcomes from a justice and well-being perspective. We argue that the results of our analysis demonstrate the need for managed retreat efforts to include a focus on compensatory, transformative approaches that center justice. This is particularly true given the types of features that we found to contribute to hazard vulnerability the most: people's sense of their own flood hazard vulnerability, their experience with discrimination, and both disability and queer identity. These characteristics are associated with larger systemic, institutional and social barriers that can make those undergoing relocation more vulnerable or subject to further discrimination in the buyout selection process or wherever they relocate to. Our findings do not necessarily suggest that relocation should not take place, but instead that such initiatives should have a strong focus on justice-centered outcomes. 



\label{sec:results3}

\subsection{Conclusion} \label{sec:conclusion}

This study illuminates the complexities that lie behind an effort to understand social vulnerability and exposure to hazards in urban settings. Additionally, while current hazard vulnerability composite indices assume a priori which proxies best represent vulnerability (such as income or education), our results show that these measures need to be coupled with more novel demographic characteristics as well as measurements of self-perception – which can be elicited through surveys, interviews, and focus groups, as well as more robust community-led policy design processes.

These indices also need to move away from a one-size-fits-all approach, as our results show that the participant characteristics that lead to more hazard exposure differ across types of hazards and across cities. For instance, risk will be defined by different characteristics when looking at extreme rain versus extreme heat. Indices should also be carefully employed with the understanding that the boundaries between resilience and vulnerability are dynamic and complex. While low-income communities might have less access to tangible resources in the face of hazard events, they may also count on stronger social networks that may be crucial to withstanding some of these events. 

It is also important to note the complexity and challenges of data collection through an international survey. Despite an iterative translation process to validate survey items across contexts, cultural perceptions of the same question may differ. Additionally, despite deploying a big N survey, the dataset obtained is not always large enough to reach conclusions for every possible way of subsetting the data (e.g., not enough experience of a certain type of hazard event among residents of a particular city). Self-reported experience with extreme weather events would also ideally be validated with comparisons to recorded events – though such validation is challenging on a hyper-local level as well as across distant locations globally. Hence, in this study we advocate for a conservative approach in the interpretation of our results. 

Future work should focus on advancing a causal understanding of the socioeconomic and political mechanisms that generate the results we find in our feature analysis. Research should aim to further pinpoint the links between urban resilience and hazard vulnerability, especially in individual cities. Utilizing novel methodologies, such as our pairing of survey elicitation and a gradient-boosting framework, can then help to increase the body of comparative urban scholarship. Finally, this avenue of research can help to develop new approaches to support those most at risk and to improve institutional approaches to urban climate adaptation. 



\bibliographystyle{abbrvnat}
\bibliography{noah} 
\end{document}